# A Multimodal Deep Learning Approach for White Matter Shape Prediction in Diffusion MRI Tractography

*Yui Lo[1,2,3], Yuqian Chen[1,2], Dongnan Liu[3], Leo Zekelman[2,4], Jarrett Rushmore[5,6], Yogesh Rathi[1,2], Nikos Makris[1,5], Alexandra J. Golby[1,2], Fan Zhang[1,2], Weidong Cai[3], and Lauren J. O'Donnell[1,2,7]*

[1] Harvard Medical School, Boston, USA

[2] Brigham and Women's Hospital, Boston, USA

[3] The University of Sydney, Sydney, Australia

[4] Harvard University, Boston, USA

[5] Massachusetts General Hospital, Boston, USA

[6] Boston University, Boston, USA

[7] Harvard-MIT Health Sciences and Technology, Cambridge, USA

**Corresponding authors**: odonnell@bwh.harvard.edu and tom.cai@sydney.edu.au

**Acknowledgments**

This work is supported by the University of Sydney International Scholarship and Postgraduate Research Support Scheme.

**Conflict of Interest**

The authors declare no conflict of interest.

**Data Availability**

The HCP-YA and ABCD datasets used in this project can be downloaded through the ConnectomeDB (db.humanconnectome.org) and PPMI Study (https://www.ppmi-info.org) websites. The ORG tractography atlas is publicly available at http://dmri.slicer.org/atlases. The code to compute shape is publicly available at https://github.com/SlicerDMRI/Tract2Shape.

**IRB Statement**

The WU-Minn HCP-YA dataset is publicly available and was approved by the institutional review board of Washington University in St. Louis (IRB #201204036). The PPMI dataset is publicly available and was approved by multiple site-specific IRBs (https://www.ppmi-info.org).

# Abstract

**Introduction**: Recently, shape measures have emerged as promising descriptors of white matter tractography, offering complementary insights into anatomical variability and associations with cognitive and clinical phenotypes. However, conventional methods for computing shape measures are computationally expensive and time-consuming for large-scale datasets due to reliance on voxel-based representations.

**Methods**: To address these limitations, we introduce Tract2Shape, a novel multimodal deep learning framework that integrates geometric streamline features (as point clouds) with scalar data descriptors (as tabular data) from tractography to predict ten white matter tractography shape measures. We propose a Siamese architecture in which each subnetwork incorporates a dual-encoder design, enabling each encoder to learn modality-specific representations. To enhance model efficiency, we utilize a dimensionality reduction algorithm for the model to predict five primary shape components. The model is trained and evaluated on two independently acquired datasets: the Human Connectome Project (HCP-YA) dataset and the Parkinson's Progression Markers Initiative (PPMI) dataset. Tract2Shape is trained and tested on the HCP-YA dataset, with performance compared against state-of-the-art models. To assess robustness and generalization, we further evaluate the model on the unseen PPMI dataset.

**Results**: Tract2Shape outperforms state-of-the-art deep learning models across all ten shape measures, achieving the highest average Pearson’s *r* and the lowest normalized mean squared error (nMSE) on the HCP-YA dataset. The ablation study shows that both multimodal input and PCA benefit performance. On the unseen testing PPMI dataset, Tract2Shape maintains a high Pearson’s *r* and low nMSE, demonstrating strong generalizability in cross-dataset evaluation. In comparison with traditional voxel-representation-based shape computation, Tract2Shape achieves a 99.2% improvement in efficiency (< 0.1s per subject).

**Conclusion**: Tract2Shape enables fast, accurate, and generalizable prediction of white matter shape measures from tractography data, supporting scalable analysis across datasets. This framework lays a promising foundation for future large-scale white matter shape analysis.

**Key Points**:

1. We investigate whether deep learning could predict white matter shape measures for efficient processing and analysis with multiple modalities.

2. We apply PCA to reduce ground truth shape measures into five principal shape components that capture the most important variations.

3. We design a multimodal framework for accurate prediction using white matter point cloud representations and white matter fiber cluster tabular descriptions as complementary features.

4. We evaluate our multimodal framework on multiple datasets, showing that it outperforms state-of-the-art methods.

**Keywords**: Shape; tractography; multimodal; deep learning; white matter

# 1. Introduction

Diffusion MRI (dMRI) tractography uniquely reconstructs the brain's white matter pathways to enable the study of white matter architecture in health and disease. Traditional quantitative analyses of dMRI tractography often leverage microstructural measures, such as fractional anisotropy (FA) and mean diffusivity (MD), which quantify the directionality and magnitude of water diffusion and are sensitive to various tissue properties [Beaulieu, 2009; Jones et al., 2013]. Other traditional dMRI tractography quantitative analyses focus on measures of pathway connectivity "strength" [Zhang et al., 2022]. Although widely used, these measures do not capture the quantitative geometric characteristics of the brain's connections.

Shape measures, an alternative approach for characterizing white matter structure, have recently gained attention within the dMRI research community. For instance, previous research demonstrated that shape measures exhibit variability across the lifespan and population [Lebel et al., 2012; Schilling et al., 2023a; Schilling et al., 2023b; Yeh, 2020]. Many shape descriptors, including length, span, area, volume, and irregularity, exhibit subject-specific variability [Armocida et al., 2025; Danstrom et al., 2025; Song et al., 2015; Yoon et al., 2025]. Furthermore, our recent work demonstrated that shape features are as informative as FA, MD, and connectivity measures for the prediction of individual cognitive performance [Lo et al., 2024a; Lo et al., 2025b].

The potential of white matter shape measures has been relatively underexplored, in part due to the computational demands of conventional methods that often rely on voxel-based shape computation. MRI tractography produces geometric data in the form of sequences of 3D points, called streamlines, which can be grouped to define individual brain connections or fiber clusters that have different anatomical properties. Voxel-based shape computation methods [Schilling et al., 2023b; Yeh, 2020] require the conversion of each white matter connection to an image representation. Such an approach can be impractical when applied to large-scale dMRI tractography datasets.

Unlike voxel-based methods, geometric deep learning can offer an efficient solution for large-scale dMRI tractography analysis by leveraging the data's inherent geometric structure. Recent work has shown the potential of geometric deep learning to analyze tractography data represented using point clouds, a geometric representation of tractography data that encodes the spatial coordinates of streamline points [Astolfi et al., 2020; Chen et al., 2025; Lo et al., 2025a]. Point clouds preserve the native 3D geometry of fiber tracts, enabling models to capture detailed anatomical structures. Deep learning allows the machine to learn complex, intuitive latent spaces, capturing both structural and semantic features of 3D point clouds [Caytuiro and Sipiran, 2025; Ge et al., 2018; Qi et al., 2017]. Our preliminary study demonstrated the potential of geometric deep learning for computing white matter shape measures directly from a point cloud representation of tractography [Lo et al., 2025a].

However, geometric deep learning networks typically require fixed-size input data. In computer vision, this constraint is often addressed by randomly sampling a fixed number of points from each input point cloud [Li et al., 2018; Qi et al., 2017; Wu et al., 2019]. While this strategy effectively captures the overall geometry of the point cloud, its application to tractography data poses a significant limitation: it discards anatomically relevant summary descriptors that carry subject-specific information. These descriptors include scalar values such as the total number of points and the total number of streamlines, which are commonly stored in the file headers of several tractography data formats [Rheault et al., 06 2022; Schroeder et al., 2006] and can be considered tabular data. This tabular information is complementary to the spatial information encoded in the point coordinates but has, to date, been largely overlooked in tractography-based geometric deep learning.

In contrast to traditional unimodal deep learning, multimodal deep learning frameworks allow the integration of multiple data modalities, such as point clouds and tabular data in our context, into a unified model. This approach has gained traction in the deep learning community as a means to leverage the complementary nature of different data types to enhance model performance [Hager et al., 2023; Kristinsson et al., 2021; Xie et al., 2023; Zhang et al., 2023]. In neuroimaging, prior studies have demonstrated that multimodal fusion can enhance model accuracy by effectively capturing and addressing the distinct characteristics of each data modality [Cai et al., 2025; Luo et al., 2024; Rahaman et al., 2021; Yan et al., 2022]. Multimodal deep learning has been applied to integrate tractography with other imaging or non-imaging data sources (such as freesurfer parcellations or functional MRI data) to support tasks like white matter tract parcellation [Chen et al., 2023b; Wang et al., 2025a]. In contrast, our work focuses on a novel application: leveraging multimodal deep learning to integrate and analyze multiple data representations derived solely from a single imaging modality, namely diffusion MRI tractography.

We introduce a novel multimodal deep learning framework, *Tract2Shape*, for accurate and generalizable prediction of white matter tractography shape measures. Tract2Shape integrates two modalities, including the 3D point cloud representation of white matter tractography and the data descriptors from the raw tractography in the form of tabular data. Tract2Shape employs two parallel encoder networks that transform each modality input into a learned embedding fused for prediction. To improve learning efficiency and interpretability, we apply principal component analysis (PCA) to the ground truth shape measures, yielding five primary shape component measures that capture key geometric variations. The multimodal fusion facilitates optimal performance across independently acquired datasets. The key contributions of this work are:

1. Multimodal Shape Prediction Framework: A deep learning approach combining spatial (point cloud) and scalar (tabular) inputs for robust shape measure prediction.

2. Principal Shape Component Modeling: A dimensionality reduction strategy that extracts five informative components from ten original ground-truth shape measures.

3. Real-World Validation: A comprehensive evaluation of the model performance using both training data and unseen real-world data.

# 2. Materials and Methods

## 2.1 Overview

Our overall strategy is as follows (Figure 1). We first perform white matter parcellation to obtain fiber clusters (Section 2.2). We then calculate shape measures (10 features described below in Section 2.3). Next, we leverage the point cloud representation (Section 2.4.1) and scalar data descriptors as tabular data (Section 2.4.2) as input for the proposed multimodal deep learning framework. Finally, we train and evaluate the model, Tract2Shape (Section 2.5), to predict 10 subject-specific shape measures.

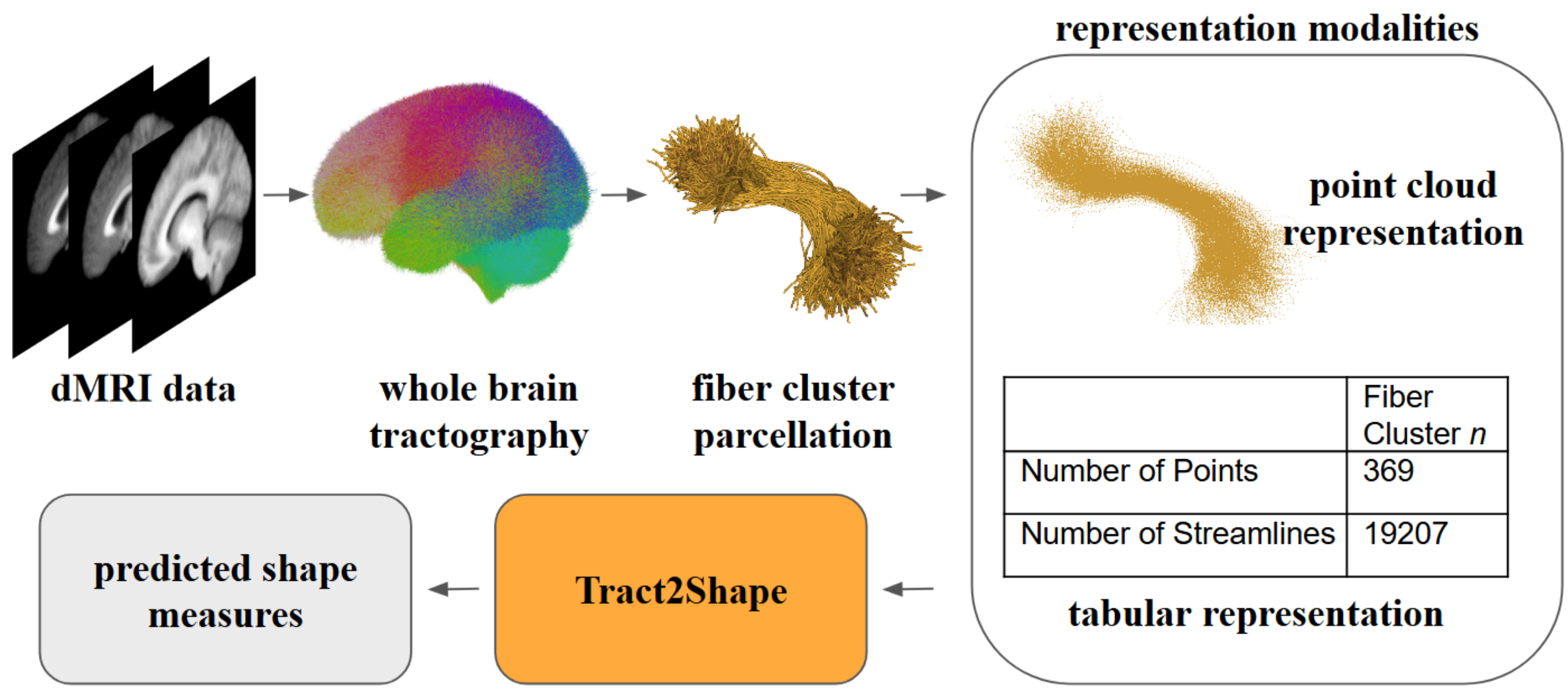

Figure 1. Overview of the proposed multimodal framework for predicting white matter tractography shape. The tractography is first parcellated into fiber clusters (top row), enabling the extraction of raw point cloud representations and the computation of tabular features such as the Number of Points and Number of Streamlines (right). These point clouds and tabular inputs are fed into the Tract2Shape model to predict shape measures represented as PCA components.

## 2.2 dMRI Tractography Datasets

To train and evaluate the effectiveness of Tract2Shape, we use dMRI data from two datasets, the Human Connectome Project minimally preprocessed young adults dataset (HCP-YA) and Parkinson's Progression Markers Initiative (PPMI). We train on the HCP-YA dataset from 1065 healthy young adults (575 females and 490 males, mean age 28.7 years) [Van Essen et al., 2012; Van Essen et al., 2013]. For evaluation, we leverage the PPMI dataset, which consists of 30 older adult subjects between the ages of 51 and 75 years (9 females and 21 males, mean age 63.1 years), including 16 Parkinson's disease (PD) patients and 14 healthy control individuals [Marek et al., 2011]. Both datasets were independently acquired to represent different populations using different imaging protocols and scanners.

Whole-brain tractography for HCP-YA and PPMI datasets [Zekelman et al., 2022; Zhang et al., 2018c] was generated for each subject's dMRI data with a multi-tensor unscented Kalman filter (UKF) tractography method [Reddy and Rathi, 2016][1] known for its consistency across the human lifespan, across test-retest scans, across disease states, and across different acquisitions [Zhang et al., 2018c; Zhang et al., 2019]. The UKF method estimates a tissue microstructure model during fiber tracking and provides tract-specific microstructural measures using the first tensor, which models the traced tract. Next, the whole-brain tractography of each subject was parcellated into fiber clusters using the whitematteranalysis package[2] to apply the O'Donnell Research Group (ORG) fiber cluster atlas [Zhang et al., 2018c]. The whitematteranalysis method parcellates tractography into fiber clusters robustly using a spectral embedding of streamlines, a machine-learning technique that considers the variability across subjects [O'Donnell and Westin, 2007]. Fiber clusters are a compact representation of the connectome with improved power to predict human traits [Liu et al., 2023; Zhang et al., 2018a], enabling a variety of downstream analyses [Chen et al., 2023c; Gabusi et al., 2024; Xue et al., 2024; Zhang et al., 2018b]. The application of the ORG atlas provided subject-specific fiber clusters along with anatomical fiber tract labels for each cluster.

In this work, we focus on the key task of shape prediction for association pathways [Yeh, 2020]. Therefore, we employ tractography within six left hemisphere association tracts of arcuate fasciculus (AF), cingulum bundle (CB), extreme capsule (EmC), inferior occipito-frontal fasciculus (IoFF), inferior longitudinal fasciculus (ILF), middle longitudinal fasciculus (MdLF), and uncinate fasciculus (UF). This results in 73 fiber clusters per HCP-YA and PPMI subject, for a total of 77,745 clusters across the 1,065 HCP-YA subjects and a total of 2,185 clusters across the 30 PPMI subjects. (This total excludes a small number of five clusters that were not identified in the PPMI dataset due to anatomical and/or scan variability [Zhang et al., 2018c].)

[1] https://github.com/pnlbwh/ukftractography

[2] https://github.com/SlicerDMRI/whitematteranalysis

## 2.3 Tractography Shape Measures

### 2.3.1 Ground Truth Measures

In this work, we select ten tractography shape measures as ground truth to provide a detailed and comprehensive analysis of the tractography [Yeh, 2020]. These measures include: length (the average length of the bundle), span (the straight-line distance between the two endpoints of the bundle), curl (the ratio of length to span), elongation (the ratio of length to diameter), diameter (the estimated width of the bundle assuming a cylinder), volume (the total volume occupied by the bundle), total surface area (the total outer surface area of the bundle), total radius of end regions (the sum of the average radius at both ends of the bundle), total surface area of end regions (the sum of the surface area at both ends of the bundle), and irregularity (a measure of how much the bundle's shape deviates from a smooth cylinder). Detailed definitions for each shape descriptor are presented and defined in [Lo et al., 2025b; Yeh, 2020]. Subject-specific tractography is computed for all subjects and all datasets with the software package DSI-Studio [Yeh, 2020][3]. For each fiber cluster, this process provides ten shape measures (Figure 2), for a total of 730 shape measures per subject.

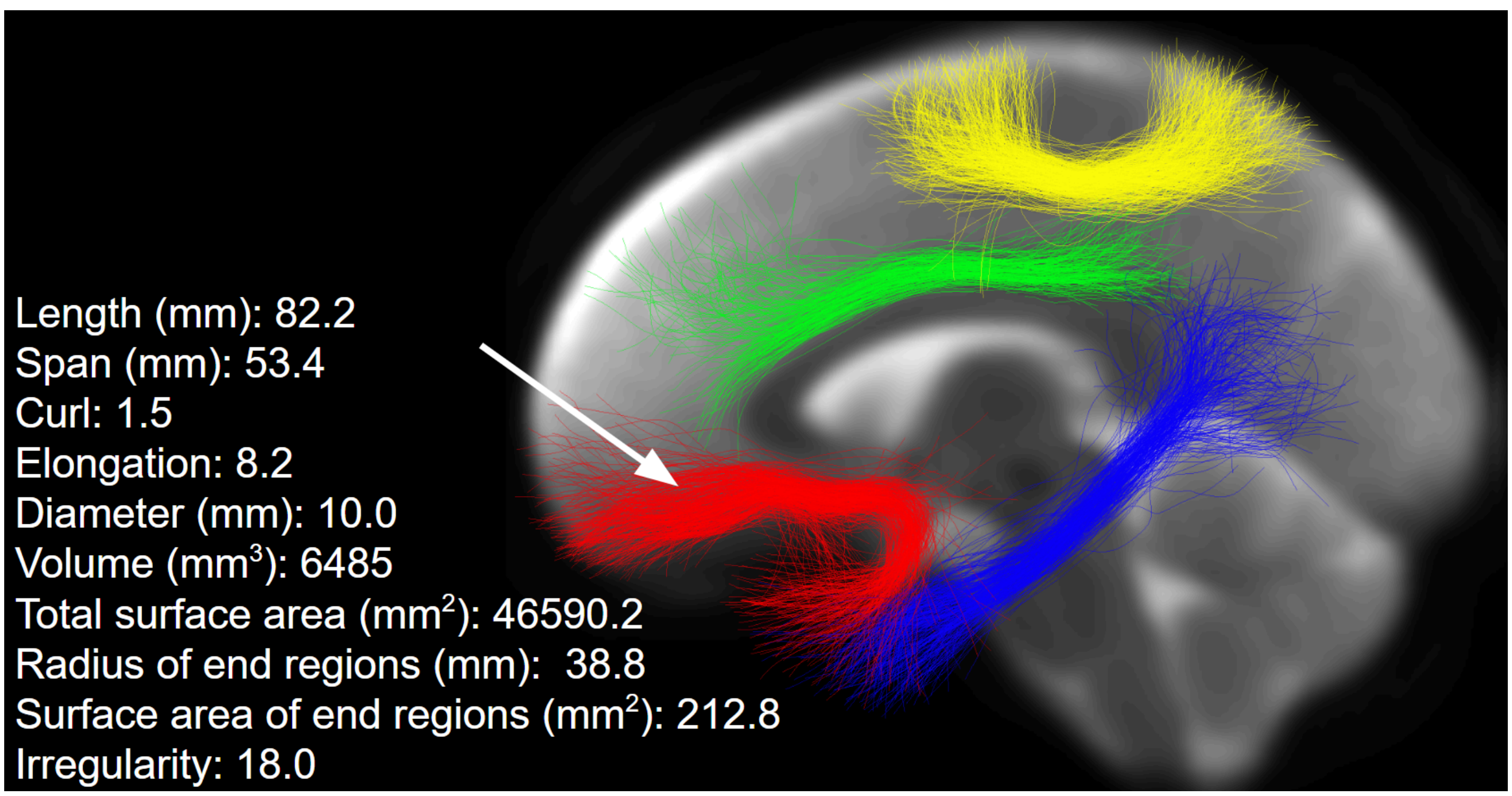

Figure 2. Four examples of individual white matter connections (fiber clusters) extracted from the entire white matter of the human brain using a fiber clustering approach. Shape measures are shown for the red fiber cluster (white arrow).

[3] https://dsi-studio.labsolver.org/

### 2.3.2 Dimensionality Reduced Shape Measures

We apply Principal Component Analysis (PCA), a widely used dimensionality reduction algorithm in brain research, to enhance computational efficiency while minimizing redundancy [Chamberland et al., 2019; Chen et al., 2023a; Migliaccio et al., 2012]. PCA projects high-dimensional data onto a lower-dimensional space while preserving the most significant variance in the dataset. We implement PCA using the Scikit-learn Python package [Pedregosa et al., 2011], which employs Singular Value Decomposition (SVD). We apply PCA to identify the principal axes of variation to capture the most meaningful shape variations, resulting in a compact representation of five dimensionality-reduced principal shape measures.

## 2.4 Multimodal Tractography Representation

### 2.4.1 Point Cloud Processing

White matter tractography is represented as point clouds to facilitate deep learning-based shape prediction. Each fiber cluster is sampled by randomly selecting N points from the tractography, a commonly employed sampling technique in point cloud processing [Chen et al., 2024; Zhang et al., 2024a; Zhang et al., 2024b]. This ensures that the geometric characteristics of the fiber cluster are preserved while reducing computational complexity. Each of the sampled points provides spatial information about the right-anterior-superior (RAS) coordinate system, a standard reference frame in neuroimaging.

### 2.4.2 Tractography Data Descriptors as Tabular Data

Geometric deep-learning networks generally require a fixed-size input, achieved by random downsampling of the input point cloud. While this sampling approach is highly successful in machine vision [Li et al., 2018; Qi et al., 2017; Wu et al., 2019] and tractography analyses [Chen et al., 2024; Zhang et al., 2024a; Zhang et al., 2024b], it discards anatomically relevant summary descriptors that encode subject-specific information about the original tractography data. In particular, our proposed multimodal tractography representation includes two scalar features per cluster: the total number of streamlines (NoS) and the total number of points (NoP). These are fundamental tractography data descriptors [Rheault et al., 06 2022; Schroeder et al., 2006] that are directly available from the raw tractography files (vtk polydata in this study) but are not available in the sampled point cloud. Therefore, these values are extracted from the original tractography files and are employed to provide complementary information along with the sampled point cloud. Both NoS and NoP are intrinsic to the tractography process, summarizing the overall geometry contained in the raw tractography files. NoS, in particular, also has neuroscientific relevance, as it often serves as a measure of white matter connectivity “strength” [Zhang et al., 2022]. By incorporating the scalars NoS and NoP as a tabular modality input feature, we provide the shape prediction model with additional information alongside the point cloud.

## 2.5 Multimodal Framework

This study introduces a hybrid network architecture, Tract2Shape, for tractography shape prediction that jointly leverages 3D point cloud representations of white matter tractography and tabular data representing associated structural properties. To effectively integrate the different data modalities to enhance prediction performance, we design a dual-encoder framework, as shown in Figure 3.

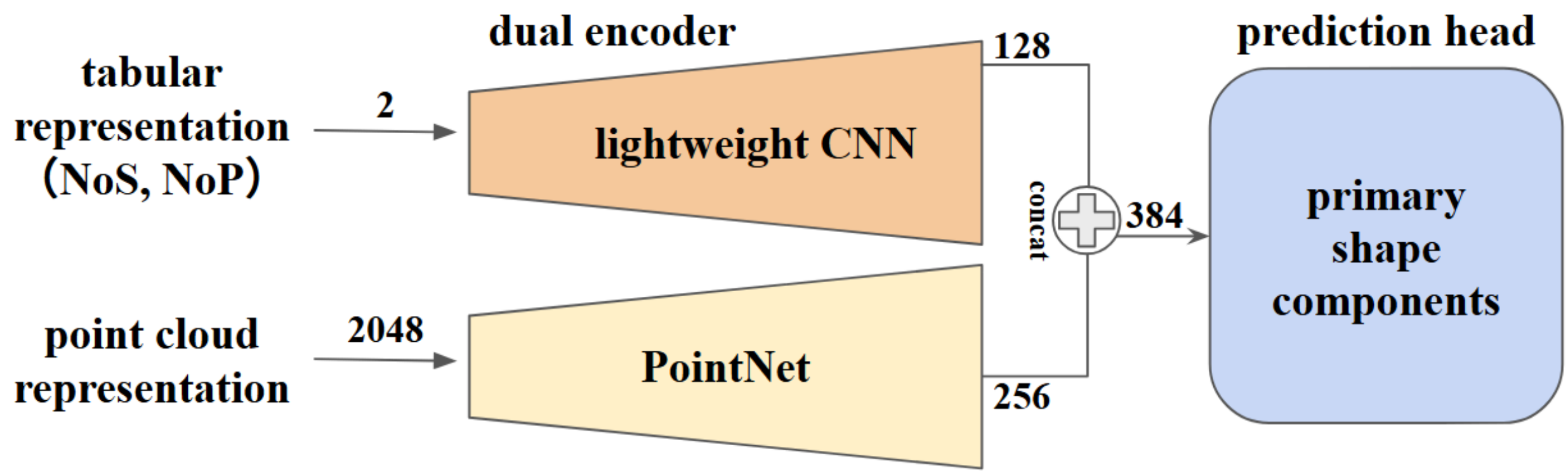

Figure 3. Overview of the proposed dual encoder for each of our subnetworks. The tabular encoder outputs a feature of size 128, and the point cloud encoder outputs a feature of size 256. These are concatenated into a feature of size 384, followed by a fully connected layer that outputs 5 PCA-reduced shape components.

In this study, we extend TractGeoNet [Chen et al., 2024], which leverages a Siamese neural network backbone and performs regression using point clouds as inputs. A Siamese network consists of two identical subnetworks with shared weights that independently process paired inputs, enabling the model to learn similarity relationships through metric learning [Bromley et al., 1993; Chicco, 2021; Lo et al., 2021]. We adopt the Siamese neural network backbone and modify each subnetwork to process the different input modalities jointly. Unlike TractGeoNet, we incorporate a dual-encoder design for each subnetwork to learn point cloud representations and tabular data representations of tractography. We adopt PointNet [Qi et al., 2017], a widely used point cloud-based neural network in tractography [Chen et al., 2025; Lo et al., 2025a; Xue et al., 2023a], as the encoder for extracting geometric embeddings from the 3D tractography point cloud representations. In parallel, we adopt a lightweight convolutional neural network (CNN) to extract data descriptor embeddings from the complementary tabular data.

The point cloud encoder produces a learned embedding of size 256, while the tabular encoder outputs a 128-dimensional feature. These embeddings are concatenated into a 384-dimensional feature vector. The concatenated embeddings from each subnetwork are passed through a fully connected layer prediction head that outputs a five-dimensional vector, representing the predicted five principal shape components. We adopt a Paired Siamese Regression Loss [Chen et al., 2024] to combine a mean prediction loss across the subnetworks and a pairwise consistency loss. Specifically, we calculate a pairwise loss based on the mean squared difference between the predicted difference vector and the actual difference in their ground truth principal shape components. After the network is trained to predict the five primary components in the dimensionality-reduced latent space, we apply the inverse PCA transformation to reconstruct the ten shape measures for model performance evaluation.

## 2.6 Implementation Details

The ground truth shape measures for the white matter tractography are computed with DSI-Studio (v.2023.07.06 "Chen" Release) [Yeh, 2020]. We optimize Tract2Shape with the Adam algorithm [Kingma and Ba, 2014] with a weight decay of 0.005. The scheduler updates the learning rate of 0.0005 with a decay factor (gamma) of 0.1 every 200 steps. Additional hyperparameters include a batch size of 128 and training for 200 epochs. All training experiments were run with the same hyperparameters to ensure reproducibility and consistency in model training and evaluation. The code is designed and implemented using PyTorch 1.13 [Paszke et al., 2019]. All experiments in this work were performed on the Jetstream2 cloud computing environment and used 10GB of GPU memory for training [Boerner et al., 2023; Hancock et al., 2021]. All training experiments were run with the same hyperparameters to ensure reproducibility and consistency in model training and evaluation. The HCP-YA dataset was randomly split cluster-wise into 80% training and 20% test sets using a fixed random seed, where each cluster was treated as an independent unit.

# 3. Experiments and Results

We train a prediction model using the HCP-YA training data from Section 2.4. To balance dimensionality reduction with information retention, we select five primary components. These five components preserve 99.2% of the total variance in the HCP-YA training dataset (Table S1) while reducing the dimensionality of the shape representation by half. The HCP-YA-derived PCA shape components represented 98.70% of the total variance in the unseen testing dataset (PPMI). Tract2Shape thus predicts five dimensionality-reduced primary component shape measures, reduced from the ten shape measures, and reconstructs back into ten shape measures.

We perform several experiments. First, we assess performance on the unseen HCP-YA test data in comparison with other methods (Section 3.2). Second, we perform an ablation study (Section 3.3). Third, we evaluate model robustness on the unseen PPMI testing dataset (Section 3.4).

## 3.1 Evaluation Metrics

Pearson’s correlation coefficient (Pearson’s *r*) [Sedgwick, 2012] and normalized mean squared error (nMSE) are employed to evaluate the performance of our proposed model and enable comparisons with the state-of-the-art approaches. Pearson’s *r* is a widely applied evaluation metric used in neuroimaging analyses [Gu et al., 2022; Kim et al., 2021; Lo et al., 2024b; Wu et al., 2023; Xiao et al., 2021]. Pearson's *r* measures the strength and direction (positive or negative) of the correlation between predicted and ground-truth shape measurements. The normalized Mean Squared Error (nMSE) metric is used in neuroimage analysis to quantify the error between predicted and ground truth values [Lemkaddem et al., 2014; Li et al., 2021; Lo et al., 2025a]. The error is normalized to a range between 0 and 1, with values near 0 indicating a closer match to the ground truth, thus, a lower prediction error.

## 3.2 Performance Evaluation and Comparative Analysis

The proposed model is compared with three state-of-the-art (SOTA) deep learning point cloud models: PointNet [Qi et al., 2017], TractGeoNet [Chen et al., 2024], and TractShapeNet [Lo et al., 2025a]. PointNet is a widely applied point-based neural network designed for handling point cloud data, and it has proven effective in tractography applications. [Chen et al., 2023b; Lo et al., 2025a; Xue et al., 2023b]. Our previous studies introduced TractGeoNet [Chen et al., 2024] to predict subject-specific language performance and TractShapeNet [Lo et al., 2025a] to predict five shape measures (length, span, volume, total surface area, and irregularity) using tractography point cloud representations. Tract2Shape extends TractGeoNet by introducing a dual-encoder architecture for each subnetwork to process multiple data modalities. Each encoder is designed to handle one modality, allowing the model to capture features from different modalities effectively.

Table 1: Comparative performance evaluation for prediction of the ten shape measures by our proposed model and state-of-the-art deep learning models trained and tested on HCP-YA. The best-performing results across the different models are highlighted in bold. All results are reported in terms of Pearson's *r*.

| Shape | PointNet | TractGeoNet | TractShapeNet | Tract2Shape |
|---|---|---|---|---|
| Length | 0.965 | 0.977 | 0.981 | **0.981** |
| Span | 0.972 | 0.982 | 0.984 | **0.985** |
| Curl | 0.909 | 0.952 | 0.955 | **0.956** |
| Elongation | 0.622 | 0.718 | 0.716 | **0.834** |
| Diameter | 0.524 | 0.608 | 0.655 | **0.915** |
| Volume | 0.711 | 0.748 | 0.766 | **0.961** |
| Total Surface Area | 0.818 | 0.847 | 0.863 | **0.953** |
| Total Radius of end regions | 0.757 | 0.834 | 0.851 | **0.862** |
| Total Surface area of end regions | 0.492 | 0.556 | 0.597 | **0.976** |
| Irregularity | 0.714 | 0.807 | 0.847 | **0.876** |
| Average | 0.748 ± 0.171 | 0.803 ± 0.148 | 0.822 ± 0.136 | **0.930 ± 0.055** |

Table 2: Comparative performance evaluation for prediction of the ten shape measures by our proposed model and state-of-the-art deep learning models trained and tested on HCP-YA. The best-performing results across the different models are highlighted in bold. All results are reported in terms of nMSE.

| Shape | PointNet | TractGeoNet | TractShapeNet | Tract2Shape |
|---|---|---|---|---|
| Length | 0.014 | 0.007 | 0.006 | **0.002** |
| Span | 0.010 | 0.006 | 0.005 | **0.003** |
| Curl | 0.124 | 0.058 | 0.053 | **0.007** |
| Elongation | 0.164 | 0.128 | 0.142 | **0.044** |
| Diameter | 0.042 | 0.036 | 0.033 | **0.009** |
| Volume | 0.116 | 0.101 | 0.094 | **0.021** |
| Total Surface Area | 0.053 | 0.044 | 0.039 | **0.015** |
| Total Radius of end regions | 0.012 | 0.009 | 0.008 | **0.007** |
| Total Surface area of end regions | 0.222 | 0.201 | 0.187 | **0.026** |
| Irregularity | 0.013 | 0.009 | 0.007 | **0.004** |
| Average | 0.077 ± 0.075 | 0.060 ± 0.065 | 0.057 ± 0.064 | **0.014 ± 0.013** |

As shown in Tables 1 and 2, our proposed model achieves the best performance across all ten shape measures, demonstrating the model's ability to learn and predict meaningful shape measures. Compared with state-of-the-art deep learning models, Tract2Shape consistently outperforms baseline models across all evaluation metrics. To assess statistical significance, we performed a one-way repeated measures ANOVA, which identified a significant performance difference across the models in Tables 1 and 2 ($p < 0.001$). Pearson's r values were first transformed using Fisher's r-to-z transformation, while nMSE values were used directly [Chen et al., 2024; Lo et al., 2025b]. Post hoc paired t-tests demonstrated that Tract2Shape significantly outperformed the compared methods (PointNet, TractGeoNet, and TractShapeNet) across both r and nMSE metrics ($p < 0.05$). These results highlight the effectiveness of integrating point cloud and tabular data, as Tract2Shape yields the lowest prediction errors and highest Pearson correlation coefficients in reconstructing the ten shape measures.

## 3.3 Computation Efficiency Evaluation

To assess efficiency, we record the computation time for each comparison method to perform tractography shape prediction in Table 3. The total runtime per subject includes both model inference and reconstruction of the original shape measures using an inverse PCA step. The results indicate that Tract2Shape and the other deep learning methods are much faster than DSI-Studio in terms of processing and computation speed.

Table 3: Comparative shape computation time evaluation (average total time per subject across 10 example subjects) between our proposed model, state-of-the-art geometric deep learning models, and the voxel-based shape computation algorithm of DSI-Studio. All times are reported in seconds.

| Method | DSI-Studio | PointNet | TractGeoNet | TractShapeNet | Tract2Shape |
|---|---|---|---|---|---|
| Total time per subject (sec) | 10.4 | < 0.1 | < 0.1 | < 0.1 | < 0.1 |

## 3.4 Ablation Study

To evaluate the effectiveness of each contribution in our proposed architecture, we conduct an ablation study using four model variants: (1) Vanilla, based on the original TractGeoNet architecture with single-modality point cloud input to directly predict all 10 original shape measures; (2) Multimodal, which extends the vanilla setup by incorporating a complementary (tabular) modality, also predicting all 10 shape measures directly; (3) PCA, where only dimensionality reduction via PCA is added to the vanilla mode by predicting the top 5 principal components, followed by inverse PCA reconstruction to recover all 10 measures; and (4) Multimodal with PCA, which integrates both multimodal input and PCA-based shape representation.

Table 4: Ablation study results for predicting ten shape measures, evaluated on HCP-YA. The best-performing results across the different models are highlighted in bold. All results are reported in terms of Pearson's *r*.

| Shape | Vanilla | Multimodal | PCA | Tract2Shape |
|---|---|---|---|---|
| Length | 0.977 | 0.978 | 0.978 | **0.981** |
| Span | 0.982 | 0.982 | 0.980 | **0.985** |
| Curl | 0.952 | 0.953 | 0.951 | **0.956** |
| Elongation | 0.718 | 0.739 | 0.747 | **0.834** |
| Diameter | 0.608 | 0.862 | 0.648 | **0.915** |
| Volume | 0.748 | 0.926 | 0.764 | **0.961** |
| Total Surface Area | 0.847 | 0.917 | 0.861 | **0.953** |
| Total Radius of end regions | 0.834 | 0.836 | 0.852 | **0.862** |
| Total Surface area of end regions | 0.556 | 0.942 | 0.585 | **0.976** |
| Irregularity | 0.807 | 0.810 | 0.857 | **0.876** |
| Average | 0.803 ± 0.148 | 0.895 ± 0.080 | 0.822 ± 0.135 | **0.930 ± 0.055** |

Table 5: Ablation study results for predicting ten shape measures, evaluated on HCP-YA. The best-performing results across the different models are highlighted in bold. All results are reported in terms of nMSE.

| Shape | Vanilla | Multimodal | PCA | Tract2Shape |
|---|---|---|---|---|
| Length | 0.007 | 0.007 | 0.003 | **0.002** |
| Span | 0.006 | 0.006 | 0.003 | **0.003** |
| Curl | 0.058 | 0.058 | 0.007 | **0.007** |
| Elongation | 0.128 | 0.120 | 0.063 | **0.044** |
| Diameter | 0.036 | 0.017 | 0.028 | **0.009** |
| Volume | 0.101 | 0.042 | 0.096 | **0.021** |
| Total Surface Area | 0.044 | 0.026 | 0.040 | **0.015** |
| Total Radius of end regions | 0.009 | 0.008 | 0.008 | **0.007** |
| Total Surface area of end regions | 0.201 | 0.059 | 0.18 | **0.026** |
| Irregularity | 0.009 | 0.008 | 0.004 | **0.004** |
| Average | 0.060 ± 0.065 | 0.035 ± 0.036 | 0.043 ± 0.057 | **0.014 ± 0.013** |

The results, summarized in Tables 4 and 5, show that each proposed improvement contributes to improved performance. The multimodal model outperforms the Vanilla baseline, demonstrating that incorporating tractography data descriptors alongside geometric point cloud data enhances the model's ability to capture complex tractography shape characteristics. Similarly, adding PCA dimensionality reduction improves predictive performance by guiding the model to learn compact, informative shape representations. Combining both components (Multimodal with PCA) yields the best results across all shape measures, confirming the complementary value of integrating heterogeneous modalities with supervised learning.

## 3.5 Model Robustness Evaluation on Unseen Data

To assess Tract2Shape's robustness and generalization, we evaluate the performance of Tract2Shape on the independently acquired PPMI testing dataset (Table 6). This cross-dataset evaluation tests the model's ability to generalize across different acquisitions, different populations, and different health conditions. The cross-dataset evaluation demonstrates a comparable performance to the in-domain performance on HCP-YA reported in Tables 4 and 5. The small performance drop in cross-dataset testing underscores the model's generalization ability and robustness across different clinical datasets in real-world applications.

Table 6: Performance evaluation for prediction of the ten shape measures. The model is trained on HCP-YA and evaluated for robustness on PPMI, an unseen dataset. All results are reported in terms of Pearson's *r* and nMSE.

| Shape | HCP-YA (train) → PPMI (unseen test) | |
|---|---|---|
| | Pearson's *r* | nMSE |
| Length | 0.906 | 0.011 |
| Span | 0.942 | 0.015 |
| Curl | 0.847 | 0.025 |
| Elongation | 0.894 | 0.065 |
| Diameter | 0.936 | 0.017 |
| Volume | 0.963 | 0.032 |
| Total Surface Area | 0.932 | 0.031 |
| Total Radius of end regions | 0.705 | 0.019 |
| Total Surface area of end regions | 0.972 | 0.043 |
| Irregularity | 0.747 | 0.016 |
| Average | 0.884 ± 0.091 | 0.027 ± 0.016 |

# 4. Discussion and Conclusion

This study presents a deep learning pipeline, Tract2Shape, for efficient and subject-specific prediction of tractography shape measures from large-scale tractography data. Our approach adopts a multimodal strategy that integrates point cloud and tabular representations of tractography to enable accurate and scalable shape estimation. To enhance model learning, we apply PCA to reduce the dimensionality of the ground truth measures. Trained on the young adult HCP-YA dataset, our framework demonstrates strong predictive performance across individual shape measurements, outperforming existing state-of-the-art deep learning models. Evaluation of the model on an unseen, independently acquired dataset acquired in older adults further demonstrates the model's generalizability and scalability. Notably, both our method and other state-of-the-art deep learning models accelerate the computational process, achieving a prediction speed faster than 0.1 seconds per subject. These results highlight the potential of our framework as a fast and effective tool for large-scale tractography shape analysis in future neuroimaging applications.

To the best of our knowledge, this is among the first deep-learning studies to predict a comprehensive set of shape measures for tractography, leveraging a multimodal approach. Our earlier study was limited to predicting five shape measures and reported suboptimal performance, particularly on geometrically complex features, such as elongation and surface area of end regions [Lo et al., 2025a]. Our Tract2Shape approach addresses this limitation by achieving consistently high accuracy across both simple and complex shape measures. Our multimodal strategy allows the model to leverage complementary tractography information and avoid reliance on a single modality.

In this work, we employed PCA to generate low-dimensional representations of dMRI features. Prior work has used PCA to reduce data from diffusion tensor imaging, NODDI, and other modalities to three components for studying white matter development [Geeraert et al., 2020]. Similarly, PCA and autoencoders have enabled improved detection of white matter anomalies from microstructure features like FA, MD, and spherical harmonic features [Chamberland et al., 2021]. In contrast, shape features from dMRI are less studied, and their dimensionality reduction remains underexplored. Dependencies among microstructure features [Ennis and Kindlmann, 2006] and among shape measures (e.g., curl derived from length and span) [Yeh, 2020] can influence PCA by amplifying correlated variables. To examine this, we performed dimensionality reduction and prediction for six primary shape measures (length, span, volume, total surface area, total end-region radius, and total end-region surface area) and then computed the remaining four shape measures (curl, elongation, diameter, and irregularity) post hoc, as they can be derived from the primary predictions [Yeh, 2020]. Although this approach yielded reasonable results, we found that including all ten shape measures in PCA yielded better prediction performance (Table S2). In Tract2Shape, we treated all measures equally and included them directly in PCA, following strategies used for dMRI microstructure features [Chamberland et al., 2021; Geeraert et al., 2020]. Future work may consider alternative strategies for dimensionality reduction of shape features, such as deep autoencoder approaches previously applied to microstructure features [Chamberland et al., 2021].

Accurate and efficient prediction of white matter shape measures can have broad implications for both clinical practice and neuroscience research. White matter shapes may serve as complementary biomarkers for detecting and monitoring neurodevelopment and neurodegeneration [Anblagan et al., 2015; Armocida et al., 2025; Bastin et al., 2013; Batchelor et al., 2006; Drakesmith et al., 2019; Im et al., 2008; Wang et al., 2025b], and understanding brain-behavior association analyses for cognition [Lo et al., 2025b]. Efficient end-to-end prediction of shape can enhance studies of large-scale datasets, enabling population studies of white matter shape variation.

Several limitations suggest important directions for future research. First, our analysis focuses on association pathways widely studied in the neuroscience community. However, future work should include a broader range of white matter tracts by employing additional tractography. Second, we evaluate our framework using the unseen PPMI dataset, which contains data from older adults, including healthy individuals and those with Parkinson's disease. Future studies should explore datasets with greater variability across the lifespan to more thoroughly assess the model's generalization performance on different age groups and health conditions. Third, we use a single tractography method and one white matter atlas. Although comparisons across different tractography techniques and atlases are beyond the scope of this study, the proposed framework is generalizable and can be retrained using alternative tractography pipelines and atlas definitions in future brain research. Fourth, additional tabular descriptors such as demographics, age, and sex could complement the point cloud representation and potentially enhance model performance. Finally, the current implementation uses lightweight encoders, which can be substituted with more advanced deep learning architectures to enhance feature extraction and representation learning.

This work addresses the challenge of predicting white matter tractography shape measures across diverse populations and datasets by introducing a robust and generalizable deep learning framework. Our proposed Tract2Shape establishes the feasibility of an efficient, scalable shape prediction technique for tractography studies and underscores the potential of our framework as a general-purpose tool for large-scale white matter shape analysis.

# 6. Supplementary

Table S1. Pearson correlation between original DSI and PCA-reconstructed shape features from the HCP-YA dataset across varying numbers of principal components (1–10). Also reported in the bottom row is the total variance retained by each number of components. The table reports the fidelity of PCA reconstruction as dimensionality increases.

| *n_components* | *1* | *2* | *3* | *4* | *5* | *6* | *7* | *8* | *9* |
|---|---|---|---|---|---|---|---|---|---|
| length | 0.846 | 0.920 | 0.989 | 0.996 | 0.998 | 0.998 | 0.998 | 1.0 | 1.0 |
| span | 0.792 | 0.940 | 0.998 | 0.998 | 0.998 | 0.999 | 0.999 | 1.0 | 1.0 |
| curl | 0.027 | 0.237 | 0.968 | 0.976 | 0.987 | 0.989 | 0.989 | 1.0 | 1.0 |
| elongation | 0.180 | 0.762 | 0.873 | 0.881 | 0.881 | 0.896 | 0.905 | 0.910 | 0.998 |
| diameter | 0.555 | 0.965 | 0.987 | 0.988 | 0.988 | 0.999 | 0.999 | 0.999 | 1.0 |
| volume | 0.848 | 0.981 | 0.982 | 0.992 | 0.993 | 0.995 | 0.999 | 0.999 | 0.999 |
| surface area | 0.923 | 0.984 | 0.987 | 0.988 | 0.992 | 0.994 | 0.999 | 0.999 | 0.999 |
| radius of end regions | 0.479 | 0.495 | 0.660 | 0.905 | 0.999 | 0.999 | 1.0 | 1.0 | 1.0 |
| area of end regions | 0.580 | 0.916 | 0.918 | 0.962 | 0.984 | 0.991 | 0.999 | 0.999 | 1.0 |
| irregularity | 0.452 | 0.644 | 0.654 | 0.945 | 0.995 | 0.996 | 0.999 | 0.999 | 0.999 |
| explained variance ratio | 0.771 | 0.896 | 0.963 | 0.987 | 0.992 | 0.996 | 0.999 | 0.999 | 0.999 |

Supplementary Experiment 1

We performed dimensionality reduction and prediction for six primary shape measures (length, span, volume, total surface area, total end-region radius, and total end-region surface area) and then computed the remaining four shape measures (curl, elongation, diameter, and irregularity) post hoc, as they can be derived from the primary predictions [Yeh, 2020].

Table S2: Performance evaluation for the prediction of ten white matter shape measures. Pearson's *r* is reported for both direct prediction and post hoc calculation of derived measures (e.g., curl, elongation, diameter, and irregularity).

| Shape | Prediction of 6 primary shape measures, then calculation of derived measures | Tract2Shape (prediction of all measures) |
|---|---|---|
| Length | **0.981** | **0.981** |
| Span | 0.983 | **0.985** |
| Curl | 0.954 (calculated) | **0.956** |
| Elongation | **0.846** (calculated) | 0.834 |
| Diameter | **0.920 (**calculated) | 0.915 |
| Volume | 0.948 | **0.961** |
| Total Surface Area | 0.938 | **0.953** |
| Total Radius of end regions | 0.858 | **0.862** |
| Total Surface area of end regions | 0.973 | **0.976** |
| Irregularity | 0.752 (calculated) | **0.876** |
| Average | 0.915 ± 0.075 | **0.930 ± 0.055** |